# Incipient Formation of the Reentrant Insulating Phase in a Dilute 2D Hole System with Strong Interactions

Richard L.J. Qiu,[1] Chieh-Wen Liu,[1] Andrew J. Woods,[2] Alessandro Serafin,[2] Jian-Sheng Xia,[2] Loren N. Pfeiffer,[3] Ken W. West,[3] and Xuan P.A. Gao[1, *]

[1] *Department of Physics, Case Western Reserve University, Cleveland, Ohio 44106, USA.*
[2] *National High Magnetic Field Laboratory and Department of Physics, University of Florida, Gainesville, Florida 32611, USA.*
[3] *Department of Electrical Engineering, Princeton University, Princeton, New Jersey 08544, USA.*

[*]Email: xuan.gao@case.edu

**Abstract**
A new reentrant insulating phase (RIP) in low magnetic fields has been reported in the literature in strongly interacting 2D carrier systems and was suggested to be related to the formation of a Wigner crystal [e.g. Qiu et al, PRL **108**, 106404 (2012)]. We have studied the transformation between the metallic liquid phase and the low field RIP in a dilute 2D hole system with large interaction parameter $r_s$ (~20-30) in GaAs quantum wells. Instead of a sharp transition, increasing density (or lowering $r_s$) drives the RIP into a state where an incipient RIP coexists with the metallic 2D hole liquid. The non-trivial temperature dependent resistivity and the in-plane magnetic field induced enhancement of the RIP highlight the competition between two phases and the essential role of spin in this mixture phase, and are consistent with the Pomeranchuk effect in a mixture of Wigner crystal and Fermi liquid.

**Main Text:**

In 2D electron systems, the ground state is expected to be an ordered quantum electronic crystal (or Wigner crystal (WC)) when the interparticle Coulomb repulsion energy is strong enough [1]. Understanding the WC formation and WC to liquid transition has been a long-standing problem. Considering the 2D WC to liquid transition a direct first order transition, Monte-Carlo simulations obtained a critical value for the interaction parameter $r_s$, the ratio between Coulomb and kinetic energy, to be ~37 [2]. Early work on the 2D WC to liquid transition by Halperin, Nelson [3] and Young [4] predicted that the raising the temperature ($T$) would cause a two-stage melting of 2D WC, first into an anisotropic fluid phase (hexatic 'liquid crystal') with quasi-long-range bond orientation order but without positional order, then a disordered isotropic fluid at high $T$. Further theoretical calculations suggested that other complex quantum phases may exist between the 2D WC and Fermi liquid, at $1 \ll r_s < 37$ [5-10]. More generally, Jamei, Kivelson and Spivak presented a theorem concluding that the direct first order 2D WC-liquid transition is prohibited in the presence of Coulomb interactions [11], thus phase separation at mesoscopic scales and new intermediate phases (electronic micro-emulsions) are unavoidable as a result of Coulomb frustration [6, 7, 11]. A conceptually simple example of intermediate phase is short-range ordered bubbles/stripes of WC coexisting with Fermi liquid which is predicted to have interesting thermodynamic and transport properties analogous to the mixture of Helium-3 solid and fluid (e.g., the Pomeranchuk effect [12] where the large spin entropy of 2D WC drives a mixture of WC and Fermi liquid to solidify upon raising temperature, in contrast to the melting of a classical solid [6, 7]). The electronic analog of Pomeranchuk effect was believed to be seen in recent experiments on magic angle twisted bilayer graphene [13, 14] where the insulator phase is a Mott-insulator instead of WC.

Despite these important theoretical progresses on the 2D quantum WC and WC-liquid transition, experimental evidences for the 2D WC-liquid transition and interaction-driven intermediate mixture phases are scarce. The classical WC in 2D was realized in electrons confined on the surface of liquid Helium [15]. 2D quantum WC has been actively sought in electron or hole carrier systems confined in semiconductor hetero-interfaces. Metal-insulator-transition (MIT) in zero magnetic field ($B$=0) in 2D systems with high $r_s$ (>10) [16] is thought to be related to the 2D WC-liquid transition physics. Although recent current-voltage characteristics data [17, 18] taken deep in the insulator phase of the MIT suggest the existence of WC, the role of WC-liquid transition is not well understood near the MIT or in the metallic regime. One main challenge in this pursuit is that to reach the clean 2D quantum WC phase in $B$=0 where $r_s$ is very large (>~37), ultra-low carrier density and extremely high sample purity are needed so Coulomb interactions dominate over kinetic and disorder potential energies. However, an intense perpendicular magnetic field can help the WC formation by forcing electrons into the same Landau levels (LLs), raising the possibility of accessing WC in 2D electrons with not so large $r_s$ (~1-10) [19]. In the extreme quantum limit where the lowest LL is partially filled (filling factor $v$<1), the reentrant insulating phase (RIP) between the $v$=1/3 or 1/5 fractional quantum Hall (FQH) liquid and the $v$=1 integer QH in GaAs/AlGaAs [20-22] and the high $B$ insulating phase at the smallest $v$ [22–25] have been attributed to the disorder pinned WC. In addition, due to the competition between the FQH liquid and a pinned WC at fractional filling factors, anomalous magneto-resistance oscillations were observed in 2D holes at ν~1/3, suggesting that FQH liquid might be mixed with the pinned WC phase [26].

In 2D Fermion systems, a lower density is expected to give rise to stronger interactions since $r_s$ is given by $r_s = 1/(a^*\sqrt{\pi p})$ with $p$ and $a^* = \hbar^2\varepsilon/(m^*e^2)$ being the carrier density and effective Bohr radius where $m^*$, e, ℏ, $\varepsilon$ are the effective mass of carriers, electron charge, reduced Planck's constant, and dielectric constant ($m^* \approx 0.3 m_e$ and $\varepsilon = 13\varepsilon_0$ for p-GaAs). More recently, RIP between the $B=0$ metallic state and the ν =1 QH fluid was found to exist in a GaAs 2D hole system with lower density and higher $r_s$ than prior 2D systems showing the RIP at fractional ν [27]. This new RIP at low $B$ was proposed to be associated with the WC phase moving from the high B, fractional ν regime to the low $B$, ν>1 regime with an increased $r_s$ value [27-30]. Elucidating the RIP to liquid transition at low $B$ in dilute 2D carrier systems can thus shine light on the connections between the various interaction driven insulating phases, $B=0$ MIT, and the WC-liquid transition physics in interacting 2D systems.

In this Letter, we report the transition from the RIP in weak magnetic fields to metallic liquid in a dilute 2D hole system (2DHS) with $r_s$ ~20-30 in high mobility GaAs quantum wells (QWs). We examine the transport properties of the low-$B$ RIP state over extended temperature and density ranges. We find that as the hole density increases, the RIP gradually evolves into a state where the metallic liquid coexists with the incipient RIP. In this intermediate mixture state, the temperature coefficient of resistance shows unusual sign changes at intermediate temperatures, which might be associated with the heating induced freezing of the WC-liquid mixture in analogy to the Pomeranchuk effect [7, 12-14]. The application of an inplane magnetic field is shown to enhance the RIP, highlighting the essential role of spin degree of freedom or entropy in the RIP and intermediate mixture phase.

Low frequency magneto-transport measurements using lockin technique were primarily performed on four high mobility GaAs QW samples from two wafers grown on (311)A GaAs similar to previous work [27]. Unless specifically noted, the sample studied has rectangle Hall bar shape with length ~8-9mm and width ~2-3mm and the measurement current was applied along the high mobility direction $[\bar{2}33]$. The samples have $Al_{0.1}Ga_{0.9}As$ barriers and Si delta doping layers placed symmetrically at a distance of 195 nm away from the 10 nm thick GaAs QW. Back gate was placed approximately 150-300 μm underneath the sample and used to tune the hole density. All the samples have density $p \sim 1.3\times10^{10}$/cm$^2$ without gating. Sample #1, 2 have mobility µ=5×10$^5$ cm$^2$/Vs and sample #3, 4 have µ=2×10$^5$ cm$^2$/Vs without gating. Similar data were obtained in independent experiments covering temperature from $T$=1K down to $T$=10 mK in three different dilution refrigerators.

Figure 1(a) shows a qualitative sketch of the 2D WC-liquid transition phase diagram vs. the hole density ($p$) and perpendicular $B$ obtained by quantum Monte-Carlo simulations in the clean limit [31]. Taking a $B$ scan at a fixed $p$, if $p$ is high (or $r_s$ is low), one expects to observe transitions from low $B$ liquid to the WC (or RIP) at 1/3<ν<1 followed by the ν=1/3 FQH liquid and then another insulating phase due to Wigner crystallization at the highest $B$. This scenario has been the subject of extensive study [19-23] and is confirmed in our own magneto-resistivity measurements as shown in the top panel of Fig.1B for $p$=2.25×10$^{10}$/cm$^2$ in sample #1. As $p$ decreases (or $r_s$ increases), increased Coulomb interactions stabilize the WC state at lower fields. As a result, at lower $p$ and with $r_s$ approaching the critical value of zero field Wigner crystallization, not only the whole high field ν<1 regime will be dominated by the RIP or WC, the WC is also expected to emerge between the $B$=0 liquid and the ν=1 QH liquid state, as represented by the $p$ = 0.86×10$^{10}$/cm$^2$ data showing a RIP at $B$~0.22T in the bottom panel of Fig.1(b). Note that although this phase diagram ignores the disorder and assumes a direct first order transition, the qualitative agreement in the trend of WC-liquid phase boundary between theory and the density tuned RIP moving from ν<1 to ν>1 as observed in experiment corroborates that RIPs observed here are driven by the WC formation and the sample is suitable for the study of WC-liquid transition.

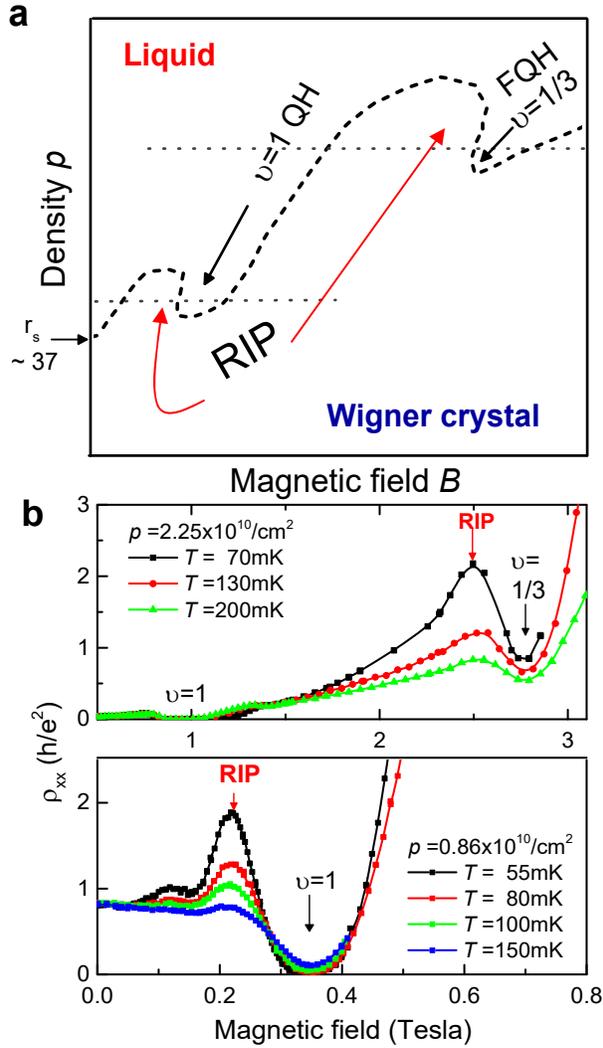

**FIG. 1.** (a) Schematic phase diagram of 2D Wigner crystal-liquid transition in the clean limit. Two horizontal lines represent the corresponding scenarios for the two field scans (top to bottom) in Fig.1(b). (b) Magneto-resistivity of a high mobility p-GaAs quantum well (sample #1) at hole densities $p=2.25$ and $0.86\times10^{10}/cm^2$ at temperatures from 55mK to 200mK. The reentrant insulating phase (RIP) is seen to move from $\nu <1$ to $\nu > 1$ as $p$ decreases.

We first investigate, at low $T$, how the low $B$ field RIP makes the transition to the metallic liquid phase as $p$ increases. When $p$ increases, $r_s$ decreases and weakened interaction effects drive the system towards the liquid state. Instead of a sharp transition from the RIP into a metallic liquid at a well-defined critical density, we observe that the RIP gradually diminishes and mixes with the metallic Fermi liquid. Figure 2(a) shows a series of $\rho_{xx}(B)$ traces at $T$=55mK from $p$=0.86 to 1.46×$10^{10}$/cm$^2$. At $p$=0.86×$10^{10}$/cm$^2$, the system's $r_s$~30 is not yet large enough to crystallize the system at $B$=0. With the application of a small perpendicular $B$, the RIP emerges as a resistance peak near 0.22T, which has a thermally activated insulating temperature dependence (Fig.3(a) and Ref. [27]). As $p$ increases, the RIP, a signature of WC, does not disappear immediately when the system develops low resistivity and clear Shubnikov de-Haas (SdH) oscillations characteristic of a Fermi liquid. Instead, the RIP peak in $\rho_{xx}(B)$ evolves into a spike inside the $v$=2 SdH dip at $B$~0.27T for $p$=1.25, 1.35×$10^{10}$/cm$^2$ and is eventually barely detected at $p$=1.46×$10^{10}$/cm$^2$. The persistence of softened RIP on top of the regular SdH oscillations deep inside the metallic regime in Fig.2 suggests that the system is in an intermediate state where the incipient insulator phase or WC is mixed with the metallic Fermi fluid background. Similar behavior was obtained in sample #2, 3, and 4, and measured under both Van der Pauw and Hall bar configurations measured along different current flow directions (Fig.S1, S2 in Supplemental Material [32]). Furthermore, this density tuned RIP to metallic liquid transition exists over a quite broad range of resistivity values (from <0.1h/e$^2$ to ~h/e$^2$, Fig.2(a), bottom to top). These observations lead to the important conclusion that the RIP to metallic liquid transition is not a sharp first order phase transition with a well-defined critical density and the system is in an intermediate mixture phase near the transition. It is noted that in the intermediate state, the Hall resistivity shows a reduced value (Fig.S3 in Supplemental Material [32] and also Ref. [33]), perhaps reflecting the inertial Hall response of WC component in the mixture state.

We next consider the suppression of the RIP due to thermal fluctuations. Figure 2(b) shows the $\rho_{xx}(B)$ curves from $T$=55mK to 150mK for $p$=1.25, 1.35 and 1.46×$10^{10}$/cm$^2$ to illustrate the effect of temperature on the magneto-resistance for the same density. As temperature decreases, we see that the incipient RIP spike near 0.27T (marked by the red arrow) grows stronger, despite the overall metallic trend in the vicinity of $v$=2 QH and very low resistivity values (<0.1h/e$^2$). This shows that the two phases (RIP and metallic QH liquid) coexist and compete. It is also noted that the width of the incipient RIP spike remains roughly the same as $T$ increases.

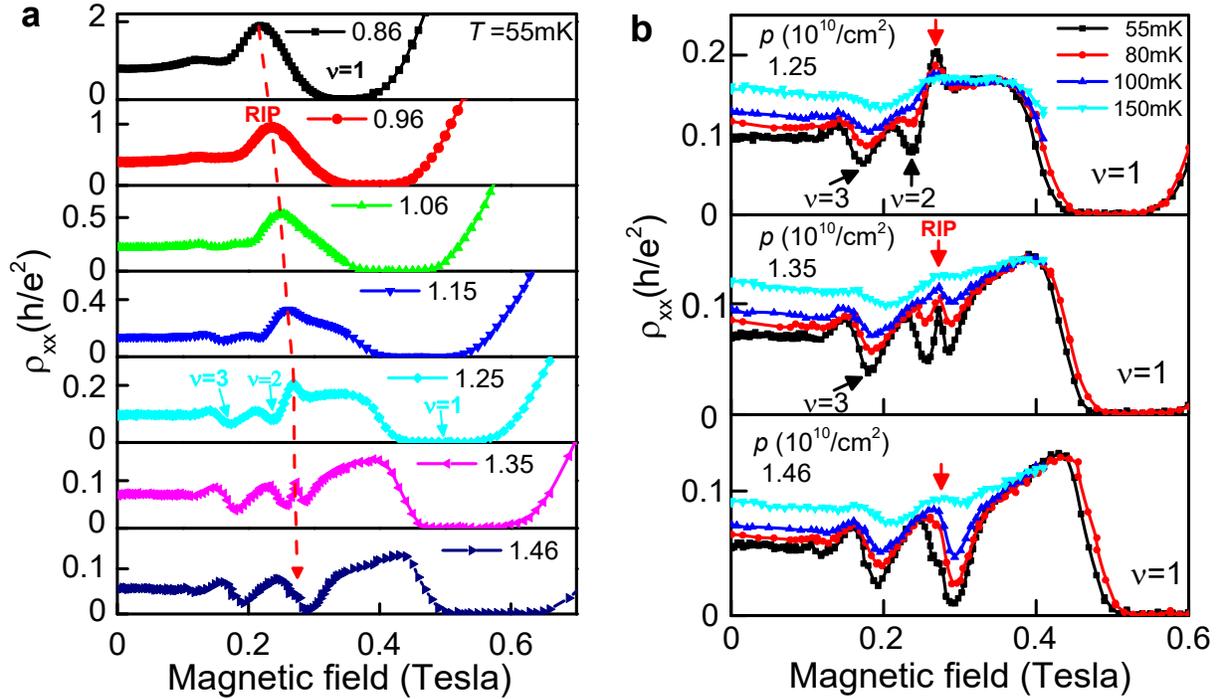

**FIG. 2.** (a) Magneto-resistivity of QW sample #1 at $T$=55mK for densities $p$= 0.86-1.46×$10^{10}$/cm$^2$ ($r_s$=30-21). The dash line illustrates the evolution of the reentrant insulating phase (RIP) or Wigner crystal while increasing the density $p$. (b) Magneto-resistivity of sample#1 for metallic densities $p$= 1.25, 1.35 and 1.46×$10^{10}$/cm$^2$ at temperatures 55mK, 80mK, 100mK and 150mK. An incipient RIP is seen at ~0.27T (marked by a red arrow).

In the WC picture of the RIP, the temperature effect corresponds to the thermal melting of the crystal. It is anticipated that $\rho(T)$ of the WC at low bias voltage follows the thermal activation law $\rho(T) \propto \exp(\Delta/2k_BT)$, with the activation energy $\Delta$ representing the energy required to create a vacancy or defect in the WC. Thermally activated temperature dependence was indeed found in earlier study in the well-developed RIP phase [27]. We now examine the temperature dependence of the incipient RIP state in more details at higher densities and to higher temperatures. The maximal resistivity at the peak of RIP, $\rho_{peak}$, is plotted in log scale against $1/T$ in Fig.3(a) for densities shown in Fig.2(a). The corresponding $\rho_{peak}$ vs. $T$ in linear scale is shown as Fig.3(b). As reported before [27], when the carrier density is low (see, e.g., $p$ = 0.86 and 0.96×$10^{10}$/cm$^2$ data in Fig.3(a)), $\rho_{peak}(T)$ indeed follows the thermal activation law at the lowest temperatures, indicated by the dashed lines. However, as $T$ increases, there is a fairly well-defined temperature $T_1$ above which the thermal activation ceases. This feature in transport was understood as the thermal melting of the 2D WC in a prior work at filling factor ν<1/5 [34]. We determine $T_1 \approx 0.3$K at $p = 0.86 \times 10^{10}$/cm$^2$ and find it decreasing with increasing $p$. At $p=1.35\times10^{10}$/cm$^2$, in spite of the existence of RIP spike in $\rho_{xx}(B)$ (Fig.2(a)), $T_1$ cannot be detected down to 55mK, the base temperature of this cool down, showing that the system remains to be a metallic liquid percolating through the incipient RIP phase down to at least 55mK.

At temperatures above $T_1$, the system shows an interesting and intriguing behavior in the incipient RIP state. When the hole density is low ($p = 0.86$ and $0.96\times10^{10}$/cm$^2$), the RIP's temperature dependent resistance changes from a rapidly increasing exponentially function into a more slowly changing insulating-like $T$-dependence at $T>T_1$, (which can be more clearly seen in Fig.3(b)). This insulating-like T-dependence at high $T$ corresponds to the transport characteristics of a melted WC in the WC interpretation of RIP. However, as the hole density increases, the starting temperature of this slowly changing insulating-like $\rho(T)$ in the high $T$ regime does not continue to coincide with $T_1$, the starting temperature of the thermally activated insulating state at the lowest $T$. Instead, it moves to a higher temperature, $T_0$, as $p$ increases, in contrast to $T_1$ which decreases with increasing $p$. And there is a metallic $\rho(T)$ at intermediate temperature $T_0 > T > T_1$. We show a zoomed in plot of incipient RIP's temperature-dependent resistivity, $\rho_{peak}(T)$ in Fig.3(c) to highlight such behavior. Since the transport of QH liquid is the obvious origin of the system's metallicity at $T_0 > T > T_1$, the splitting of the cross-over temperature in the thermal melting of RIP into two characteristic temperatures with metallic-like behavior between, is quite illuminating. It shows that the system in the intermediate state of metallic QH liquid mixed with incipient RIP shows an unusual competition between the two phases: the transport is dominated by the RIP (or WC) at low and high temperatures and the metallic QH liquid dominates at intermediate temperatures. In Fig3(d), we plot both characteristic cross-over temperatures as a function of carrier density for both sample #1 and #3 [35]. In the intermediate microemulsion theory of the mixture phase between WC and Fermi liquid, an analog of Pomeranchuk effect is anticipated which may give rise to such kind of behavior: at intermediate temperatures between the Fermi temperature of the fluid and the spin ordering temperature of the WC (which is very low), raising the temperature would cause the fluid with low spin entropy to freeze into spin disordered WC and the system's resistance increase [7]. The increasing trend of $T_0$ with $p$ (and thus the Fermi temperature) and the decreasing trend of $T_1$ with $p$ appears to be qualitatively consistent with the Pomeranchuk interpretation.

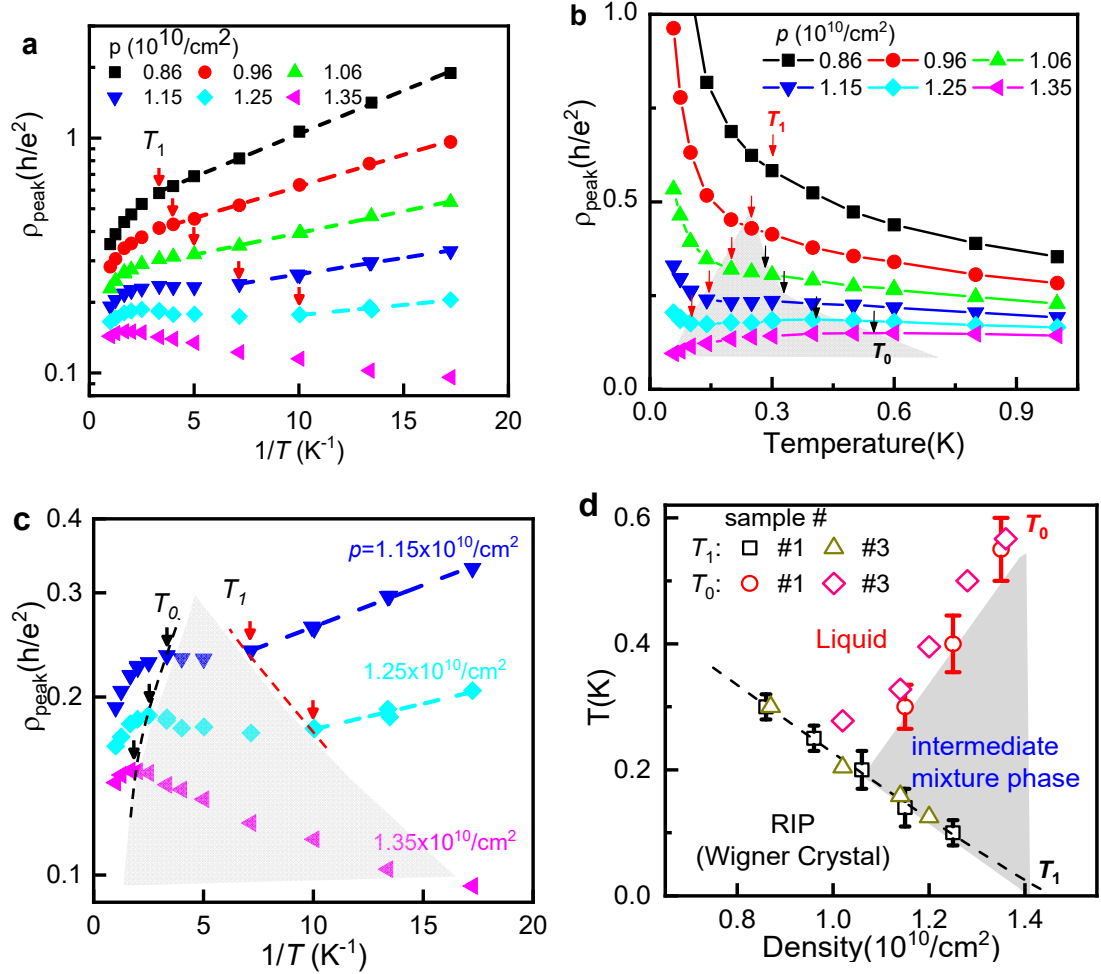

**Fig. 3.** (a) Temperature dependence of resistivity at the peak of RIP from QW sample #1 plotted vs. $1/T$ scale at densities $p = 0.86, 0.96, 1.06, 1.15, 1.25$, and $1.35 \times 10^{10}/cm^2$ (corresponding magnetic fields are 0.220, 0.235, 0.250, 0.260, 0.268, 0.272, and 0.274Tesla). The dash lines are the fits to activation behavior $\rho \propto \exp(\Delta/2T)$. $T_1$ marks the point when activated $\rho(T)$ terminates. (b) Data in (a) plotted in linear scale. (c) At the densities where an incipient RIP coexists with 2D QH fluid, the system shows insulating behavior at both low ($<T_1$) and high ($>T_0$) temperatures with a metallic behavior in the intermediate temperature range ($T_0>T>T_1$). (d) The two characteristic temperatures $T_0$ and $T_1$ plotted vs. 2D hole density for both sample #1 and #3.

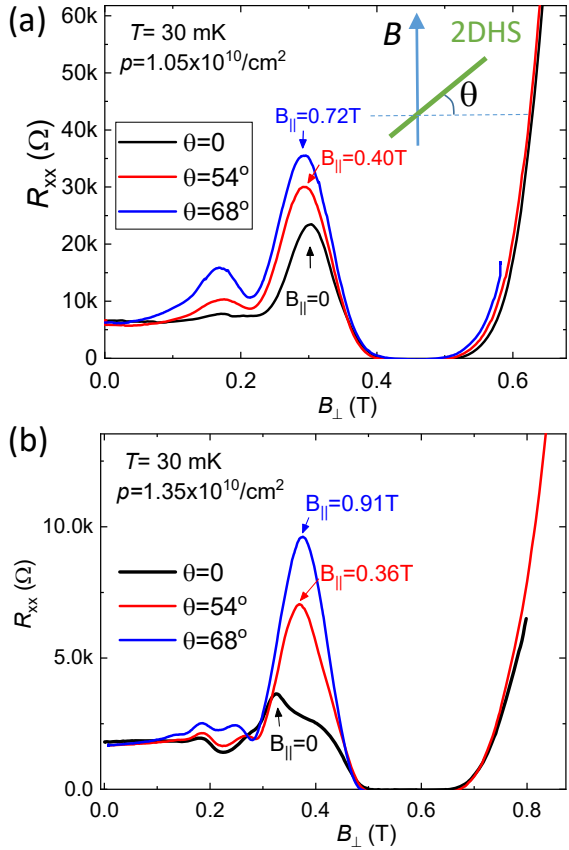

**Fig. 4.** Magneto-resistance of a dilute 2DHS (sample #3) with density $p=1.05\times10^{10}/cm^2$ (a) and $1.35\times10^{10}/cm^2$ (b) in magnetic field at a tilt angle $\theta$ (schematic shown in the inset of (a)) at $T=30mK$. The arrows mark the RIP peak and the corresponding inplane field component $B_\parallel$ at the RIP peak.

Applying an inplane magnetic field to the 2D systems is a powerful means to probe the spin effect via polarizing the spins of the system. In 2D MIT at $B=0$, the application of an inplane magnetic field greatly increases the resistivity of the system and quenches the metallic behavior [16, 36]. We further investigated the effect of spin-polarization on the RIP. A shown in Fig.4 (a) inset, the sample is placed in magnetic field at a fixed tilt angle $\theta$ and the magnetic field is swept while magneto-resistance is measured. In this configuration, both the perpendicular field component $B\perp$ and inplane field component $B_\parallel$ increase with the magnetic field according to $B\cos(\theta)$ and $B\sin(\theta)$. Fig.4(a) shows the results for $p=1.05\times10^{10}/cm^2$ where the RIP is relatively well formed (at ~ 0.3T perpendicular $B$) and Fig.4(b) shows the results for $1.35\times10^{10}/cm^2$ where a weak incipient RIP is seen at ~ 0.33T perpendicular $B$ at $\theta=0$. When the sample is tilted, increasing the $B$ sets an increasing inplane field $B_\parallel$. As a result, one sees that the RIP is enhanced. Interestingly, we can also see that while the inplane field component caused the RIP peak resistance to increase ~50% at $\theta=68°$, its effect is much more dramatic in the $p=1.35\times10^{10}/cm^2$ traces where the weak incipient RIP spike develops into a much stronger peak and the peak resistance is increased by ~300% at tilt angle $\theta=68°$. This intriguing behavior suggests that quenching the spin degree of freedom favors the RIP, similar to the zero magnetic field MIT [16] and the spin degree of freedom plays a key role in the competition between RIP and metallic QH liquid in the intermediate phase where RIP is mixed with QH liquid. In the intermediate state theory of WC and Fermi liquid mixture [7], at intermediate temperatures $T<T_F$, the Fermi temperature of Fermi liquid, due to the larger spin entropy of WC compared to the Fermi liquid, WC is favored as the temperature is raised, giving rise to the Pomeranchuk effect. Due to the large spin susceptibility of WC, applying an inplane magnetic field induces a large increase in the WC's magnetization and lowers the free energy, thus the WC is favored in an inplane magnetic field. Because WC is the likely origin of the RIP, such spin polarization induced Pomeranchuk effect may be responsible for the rapid enhancement of RIP when an inplane field is applied in our experiments.

In conclusion, Coulomb interaction driven many-body phases and phase separation have been a prominent subject in many strongly correlated materials. We found evidence for the formation of an intermediate mixture phase when the low magnetic field reentrant insulating phase is suppressed by either increasing the carrier density or temperature, in a strongly interacting 2D hole system. The application of an inplane magnetic field, however, enhances the reentrant insulating phase. These observations may be related to the Pomeranchuk physics in the mixture of 2D WC and Fermi fluid and are expected to shed light on various topics such as the MIT in strongly interacting 2D electron systems and 2D Wigner crystal melting.

**Acknowledgments** X. P. A. G. thanks NSF for funding support (DMR-1607631). Measurements at the NHMFL High B/T Facility were supported by NSF grant DMR-1644779, by the State of Florida. The work at Princeton was partially funded by the Gordon and Betty Moore Foundation and the NSF MRSEC Program through the Princeton Center for Complex Materials (DMR-0819860).

# Supplementary Materials for

## Incipient Formation of the Reentrant Insulating Phase in a Dilute 2D Hole System with Strong Interactions

Richard L.J. Qiu,[1] Chieh-Wen Liu,[1] Andrew J. Woods,[2] Alessandro Serafin,[2] Jian-Sheng Xia,[2] Loren N. Pfeiffer,[3] Ken W. West,[3] and Xuan P.A. Gao[1, *]

[1]*Department of Physics, Case Western Reserve University, Cleveland, Ohio 44106, USA.*
[2] *National High Magnetic Field Laboratory and Department of Physics, University of Florida, Gainesville, Florida 32611, USA.*
[3]*Department of Electrical Engineering, Princeton University, Princeton, New Jersey 08544, USA.*

*Email: xuan.gao@case.edu

**This file includes:**

    Figs. S1 to S3

**Supplementary Figures**

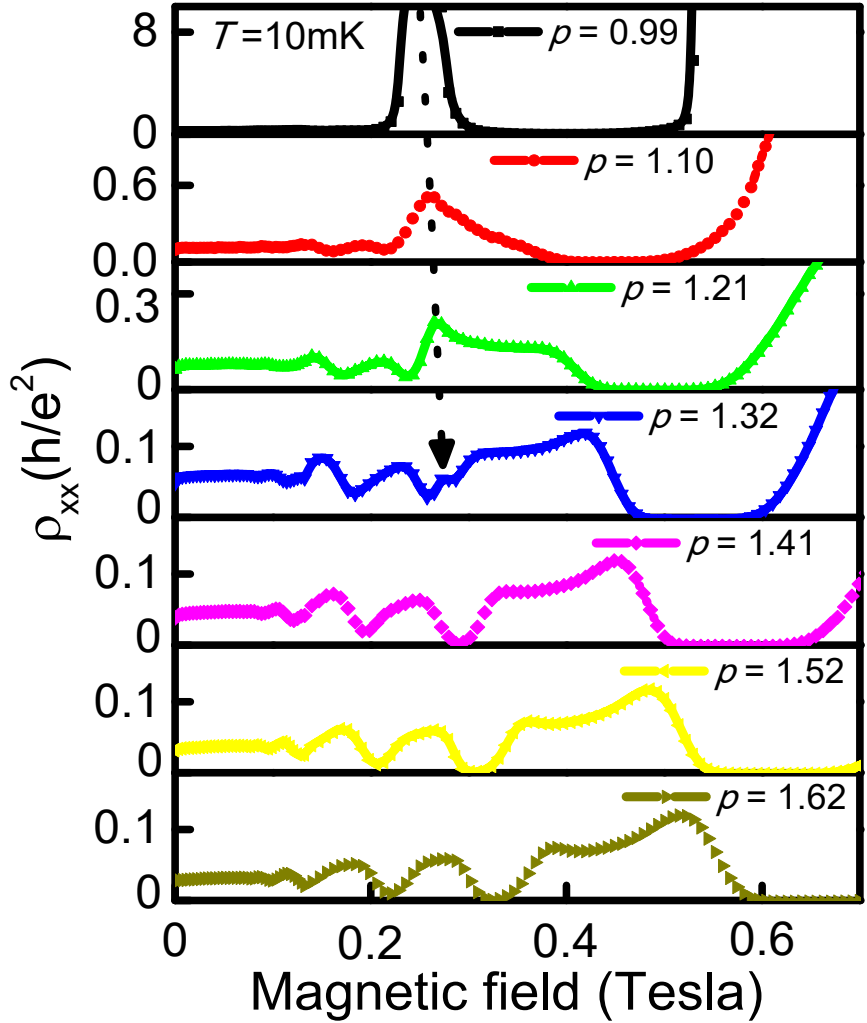

**Fig. S1** Magnetoresistivity of rectangular shaped Hall bar shaped GaAs QW (sample # 2) which has mobility $5\times10^5$ cm$^2$/Vs without gating. $T=10$mK and density $p = 0.99, 1.10, 1.21, 1.32, 1.41, 1.52$, and $1.62$ from top to bottom. The dash line illustrates the evolution/melting of the reentrant insulating phase (RIP) or 2D Wigner crystal while increasing the density $p$. For density $p = 0.99$, the RIP has peak resistivity $\rho_{peak}$ two order of magnitude larger than the resistivity $\rho(B=0T)$ at zero field.

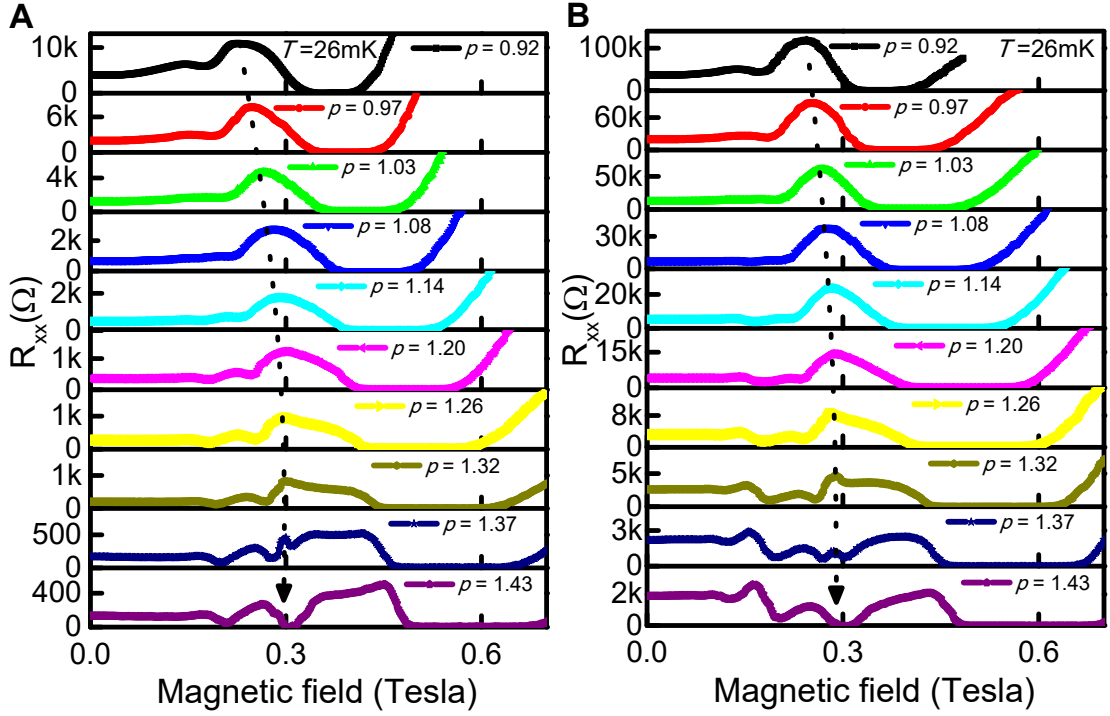

**Fig. S2 (A)** Series of magneto-resistivity plots of square shaped GaAs QW (sample #3) which has mobility $2\times10^5$ cm$^2$/Vs without gating at 26mK for hole densities $p$ = 0.92, 0.97, 1.03, 1.08, 1.14, 1.20, 1.26, 1.32, 1.37, and 1.43 $\times10^{10}$ /cm$^2$ while current flows along high mobility direction [$\bar{2}$33] for the wafer's growth orientation (311)A. The dash line illustrates the evolution/melting of the reentrant insulating phase (RIP) or 2D Wigner crystal while increasing the density $p$. **(B)** Magneto-resistivity plots as in (A) while current flows along low mobility direction [01$\bar{1}$]. The dash line illustrates the evolution/melting of RIP while increasing the density $p$.

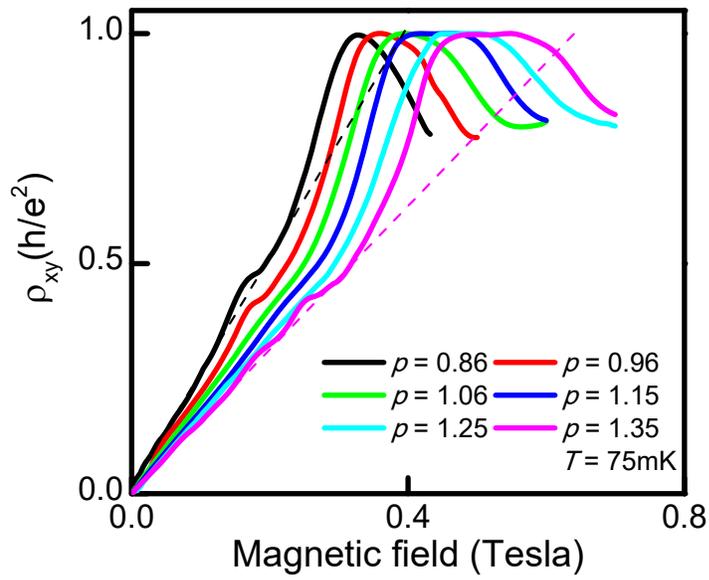

**Fig. S3.** Hall resistivity $\rho_{xy}$ vs. magnetic field $B$ for rectangular shaped Hall bar sample (#1) at $T$ = 75mK for densities $p$ = 0.86, 0.96, 1.06, 1.15, 1.25 and 1.35. The dashed lines show linear extrapolation of low field Hall resistivity.